\documentclass[12pt]{spieman}  

\usepackage{amsmath,amsfonts,amssymb}
\usepackage{graphicx}
\graphicspath{ {./figures/} }
\usepackage[colorlinks=true, allcolors=blue]{hyperref}
\usepackage{float}
\usepackage{gensymb}
\usepackage[noabbrev]{cleveref}
\usepackage{setspace}
\usepackage{tocloft}
\usepackage[left]{lineno}

\title{Phasing the Giant Magellan Telescope: Lab Experiments and First On-sky Demonstration}

\author[a,b]{Maggie Y. Kautz}
\author[b,c]{Sebastiaan Y. Haffert}
\author[b]{Laird M. Close}
\author[b]{Jared R. Males}
\author[a,b,d,e]{Olivier Guyon}
\author[f]{Alexander D. Hedglen}
\author[b]{Victor Gasho}
\author[g]{Richard Demers}
\author[h]{Antonin Bouchez}
\author[g]{Fernando Quirós-Pacheco}
\author[i] {Cédric Plantet}
\author[j]{Avalon L. McLeod}
\author[a]{Jay K. Kueny}
\author[b]{Jialin Li}
\author[a]{Joshua Liberman}
\author[b,k]{Joseph D. Long}
\author[a]{Jennifer Lumbres}
\author[a]{Eden A. McEwen}
\author[l]{Logan A. Pearce}
\author[m]{Lauren Schatz}
\author[g]{Patricio Schurter}
\author[n]{Breann Sitarski}
\author[a]{Katie Twitchell}
\author[b]{Kyle Van Gorkom}
\affil[a]{James C. Wyant College of Optical Sciences, University of Arizona, 1630 E University Blvd, Tucson, AZ, USA 85721}
\affil[b]{Steward Observatory, University of Arizona, 933 N Cherry Ave, Tucson, AZ, USA 85719}
\affil[c]{Leiden University, Rapenburg 70, 2311 EZ Leiden, Netherlands}
\affil[d]{Subaru Telescope, National Observatory of Japan, National Institutes of Natural Sciences, 650 N. A'ohoku Place, Hilo, HI, USA 96720}
\affil[e]{Astrobiology Center, National Institutes of Natural Sciences, 2-21-1 Osawa, Mitaka, Tokyo, Japan}
\affil[f]{Northrop Grumman Corporation, 600 S Hicks Rd, Rolling Meadows, IL, USA 60008}
\affil[g]{Giant Magellan Telescope Organization, 300 N Lake Ave 14th Floor, Pasadena, CA, USA 91101}
\affil[h]{W. M. Keck Observatory, 65-1120 Mamalahoa Hwy, Kamuela, HI, USA 96743}
\affil[i]{Osservatorio Astrofisico di Arcetri, Largo Enrico Fermi, 5, 50125 Firenze, Florence, Italy}
\affil[j]{Draper Laboratory, 555 Technology Square, Cambridge, MA, USA 02139}
\affil[k]{Flatiron Institute, the Center for Computational Astrophysics, 162 5th Ave 9th floor, New York, NY, USA 10010}
\affil[l]{Department of Astronomy, University of Michigan, 323 West Hall, 1085 S University Ave, Ann Arbor, MI 48109, USA}
\affil[m]{Starfire Optical Range, Kirtland Air Force Base, Albuquerque, NM, USA 87105}
\affil[n]{NASA Goddard Space Flight Center, 8800 Greenbelt Rd, Greenbelt, MD, USA 20771}

\newcommand{\edited}[1]{#1}
\newcommand{\removed}[1]{}

\cftpagenumbersoff{figure}
\cftpagenumbersoff{table} 
\begin{document} 
\maketitle

\begin{abstract}
The large apertures of the upcoming generation of Giant Segmented Mirror Telescopes (GSMTs) will enable unprecedented angular resolutions that scale as $\propto$ $\lambda$/D and higher sensitivities that scale as $D^4$ for point sources \edited{corrected by adaptive optics (AO)}. However, all will have pupil segmentation caused by mechanical struts holding up the secondary mirror [European Extremely Large Telescope (E-ELT) and Thirty Meter Telescope (TMT)] or intrinsically, by design, as in the Giant Magellan Telescope (GMT). These gaps will be separated by more than a typical atmospheric coherence length (Fried Parameter). \edited{The pupil fragmentation at scales larger than the typical atmospheric coherence length, combined with wavefront sensors with weak or ambiguous sensitivity to differential piston, can introduce differential piston areas of the wavefront known as ``petal modes".} Commonly used wavefront sensors, such as a pyramid wavefront sensor (PyWFS), \edited{also} struggle with phase wrapping caused by $>\lambda/2$ differential piston wavefront error (WFE). We have developed the holographic dispersed fringe sensor (HDFS), a single pupil-plane optic that employs holography to interfere the dispersed \edited{light from each} segment onto different spatial locations in the focal plane to sense and correct differential piston between the segments. This allows for a very high \edited{and} linear dynamic piston sensing range of approximately $\pm$10 $\mu$m. We have begun the initial attempts at phasing a segmented pupil utilizing the HDFS on the High Contrast Adaptive optics phasing Testbed (HCAT) and the Extreme Magellan Adaptive Optics instrument (MagAO-X) at the University of Arizona. Additionally, we have demonstrated use of the HDFS as a \edited{differential} piston sensor on-sky for the first time. We were able to phase each segment to within $\pm$$\lambda$/11.3 residual piston WFE (at $\lambda$ = 800 nm) of a reference segment and achieved $\sim$50 nm RMS residual piston WFE across the aperture in poor seeing conditions.
\end{abstract}


{\noindent \footnotesize\textbf{*}Maggie Y. Kautz, \linkable{maggiekautz@arizona.edu} }

\begin{spacing}{2}   

\section{INTRODUCTION}
\label{sec:intro}
All of the next generation Giant Segmented Mirror Telescopes (GSMTs) will have pupil segmentation induced by mechanical struts holding up the secondary mirror [European Extremely Large Telescope (E-ELT, \edited{25 cm struts}) and Thirty Meter Telescope (TMT, 22.5 cm struts)] or intrinsically by design as in the Giant Magellan Telescope (GMT, $\sim$30-100 cm gaps)\cite{bertrou-cantou_petalometry_2020, leboulleux_redundant_2022, janin-potiron_advancements_2024, usuda_preliminary_2014, bernstein_overview_2014, sitarski_gmt_2022}. These gaps in the pupil \edited{allow} ``petal modes", \edited{wavefront} inconsistencies between the large segments, that can introduce differential segment piston into the system. The GMT will employ free space edge-sensors to keep the primary mirrors physically in phase with each other\cite{sitarski_gmt_2022}. However, atmospheric turbulence will induce differential segment piston errors on the order of $\pm$5 $\mu$m\cite{schwartz_sensing_2017}. The large gaps between the GMT's \edited{seven} primary mirror segments are $>$10-20 cm, the typical atmospheric coherence lengths, $r_0$, at visible wavelengths. This creates a challenge for adaptive optics (AO) systems because the gaps cause missing wavefront sensor information which make it difficult to estimate piston. The AO system could settle on a different piston value for two adjacent segments and initialize a build-up of additional differential piston error\cite{schwartz_sensing_2017,bertrou-cantou_confusion_2022, hedglen_lab_2022}.

Conventional wavefront sensors in adaptive optics systems are the \edited{classical} Shack-Hartmann wavefront sensor (SHWFS) and the pyramid wavefront sensor (PyWFS), both of which \edited{struggle to} measure differential piston. The distance between segments is larger the atmospheric coherence length which also makes the wavefront over the segments uncorrelated\cite{sauvage_low_2015,sauvage_tackling_2016}. While pyramid wavefront sensors can measure differential piston, their sensitivity depends strongly on modulation radius\cite{esposito_cophasing_2003}. An unmodulated PyWFS can sense differential piston, but, like an interferometer\cite{verinaud_nature_2004}, it has phase wrapping issues and can only sense the piston up to a multiple of $\lambda$/2\cite{pinna_phase_2006, hedglen_lab_2022}. In practice, the PyWFS is never used on-sky unmodulated due to its limited dynamic range. The response of a modulated PyWFS is similar to a slope, meaning it loses sensitivity to differential piston\cite{bertrou-cantou_confusion_2022}. 

The GMT has selected a two-\edited{stage} phasing system comprised of a holographic dispersed fringe sensor (HDFS)\cite{haffert_phasing_2022} to drive the differential piston to within $\pm\lambda$/2, then a PyWFS to complete final fine phasing\cite{quiros-pacheco_giant_2022}. The HDFS uses a single pupil-plane
hologram to interfere the \edited{dispersed light from each segment} onto different spatial locations in the focal plane. Interference between \edited{light from two given} segments creates a fringe from which \edited{the amount of} differential piston can be derived\cite{chanan_phasing_1998, chanan_phasing_2000}. When the \edited{amount of differential} piston changes, the fringe pattern moves. The phase wrapping ambiguity\edited{, which also plagues a monochromatic fringe pattern,} is solved by dispersing the fringe in wavelength.

\edited{The High Contrast Adaptive optics phasing Testbed (HCAT) has been developed to demonstrate co-phasing of the seven segments of the GMT, utilizing the two-channel strategy that combines the HDFS for coarse phasing with a PyWFS for fine phasing.} The GMT's phasing requirement for the Natural Guide star AO system (NGAO: PyWFS + HDFS controlling the GMT adaptive secondary) is 45 nm RMS \edited{wavefront error (WFE)}\cite{quiros-pacheco_giant_2018}. Our \edited{initial} goal in-lab/on-sky is to utilize the HDFS to correct \edited{static} differential piston to within $\pm\lambda$/2 WFE at $\lambda$ = 800 nm. We present the first lab and on-sky phasing demonstrations using the HDFS \edited{alone as the first-stage differential piston sensor}. In Section~\ref{sec: HDFS} we will explain the functionality  and manufacturing of the HDFS. In Section~\ref{sec: lab} we will describe the lab-setup of the phasing experiments, including the optical layout of the HCAT bench and \edited{how a ``parallel DM" can correct differential piston sensed by an HDFS}. In Section~\ref{sec: in_lab} we will describe our in-lab phasing results with the HDFS. In Section~\ref{sec: on_sky} we will describe our on-sky phasing results with the HDFS.

\section{Holographic Dispersed Fringe Sensor (HDFS)}
\label{sec: HDFS}

\subsection{HDFS Background}
\begin{figure} [H]
\begin{center}
\begin{tabular}{c} 
\includegraphics[width=\textwidth]{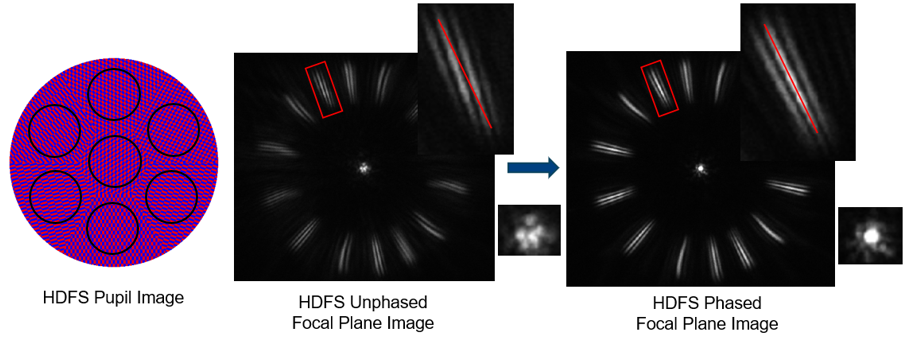}
\end{tabular}
\end{center}
\caption{The left image shows the \edited{multiplexed} diffraction gratings etched onto the HDFS optic that disperse and interfere the seven segments of the GMT pupil. The middle and right images show the focal plane image \edited{produced} by the HDFS. The segments are interfered pairwise which creates 14 dispersed interference patterns (one per segment pair for the $m = +1$ diffraction order and one for the $m = -1$ diffraction order). The 0\textsuperscript{th} order point spread function (PSF) is in the center of the pattern. The middle image shows what an unphased pupil's HDFS focal plane image would look like. Note the twist in the fringes (like a barber pole) that signifies the presence of differential piston between two segments. When there is no more differential piston in the pupil, the fringes become straight and evenly illuminated as in the rightmost image.}
\label{fig:hdfs_schem} 
    \end{figure}

The phase hologram \edited{of the HDFS} consists of several multiplexed gratings. The grating frequency determines the location of the fringes and the dynamic range of the HDFS. \edited{A larger grating frequency will create a larger spectrum and therefore decrease the spectral bandwidth per resolving element in the HDFS spectra. A smaller spectral bandwidth means the coherence length will increase which automatically translates into a larger differential piston capture range. It comes at the cost of increased detector size; a larger spectrum simply needs more pixels for adequate sampling. More pixels, however, means a larger contribution of detector noise. Therefore, there is a balance between dynamic range and sensor size and quality.} \edited{An individual grating} creates \edited{a dispersed fringe} on either side of the focal plane pattern, one for the $m = +1$ diffraction order and one for the $m = -1$. \edited{While the original experiments with the HDFS have been done with continuous gratings \cite{haffert_phasing_2022}, the gratings in this work are binary. Binary gratings proved to have higher diffraction efficiency than continuous gratings \cite{haffert_phasing_2022}.}  The highest efficiency is achieved if the amplitude of the binary grating is exactly $\pi$ radians. At that phase amplitude no light will diffract into the 0\textsuperscript{th} order and more than 80\% will end up into the $m = \pm1$ diffraction order. \edited{Due to the dispersion in the refractive index of the glass, the actual amplitude of the grating is not $\pi$ radians at all wavelengths within the spectral bandpass, causing light to leak into the 0\textsuperscript{th} order.} The rest of the light will diffract into higher diffraction orders. Another \edited{important} aspect is that binary phase masks are very easy to manufacture with photolithography. The first HDFS experiments \edited{with continuous gratings} were successful \edited{but} the manufactured HDFS optic did not completely meet the required specifications. The \edited{HDFS design in this work, shown in Figure~\ref{fig:hdfs_schem},} has a binary pattern frequency of 50 cycles/pupil \edited{and a dynamic range of $\pm$10 $\mu$m.}

\subsection{HDFS Manufacturing}
The binary HDFS is etched onto a 1" fused silica substrate. The diameter of the etched mask itself is about 9 mm which is slightly oversized compared to the \edited{8.9 mm} size of the GMT aperture at the HDFS plane. The optic was oversized to make the HDFS robust against misaligments. In attempt to maximize throughput to the focal plane, the HDFS pattern was etched onto a fused silica window with nano-textures. The nano-textures act as an extremely good anti-reflection (AR) coating \cite{wilson_optical_1982}. Depositing an AR coating on top of the HDFS might lead to unacceptable phase errors. We experimented with the nano-textured AR coatings because these are etched into fused silica and we expected that another binary etch would leave the nano-textures intact. The manufactured HDFS was investigated to determine if the second etching process damaged the AR coating. We determined this by measuring the amount of reflected light, which should be less than 1\% if the nano-textured coating was not damaged. We found that the AR coating now reflects 3.3$\%$ of the incident light, which is similar to that of uncoated glass. So, the etching process \edited{did} indeed damage the \edited{AR} coating. However, the optic still \edited{met} our desired specifications \edited{because any potential ghost would not interfere with the experiments.}
%

\section{Lab Set Up}
\label{sec: lab}
\begin{figure} [H]
\begin{center}
\begin{tabular}{c} 
\includegraphics[height=8cm]{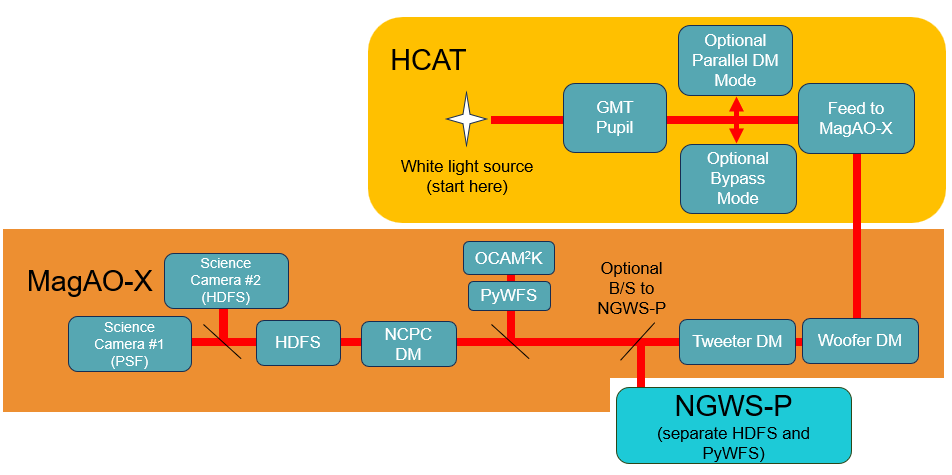}
\end{tabular}
\end{center}
\caption{A white light source on the HCAT bench propagates through the testbed and through a hole in the wall into the entrance window of MagAO-X. Segment piston can be created with piezoelectric actuators within the parallel DM, in that mode, or with the NCPC DM on MagAO-X when HCAT is in bypass mode. The HDFS is used as a differential piston sensor for either piston generator.}
\label{fig:schematic} 
    \end{figure}

\begin{figure} [H]
\begin{center}
\begin{tabular}{c} 
\includegraphics[height=8cm]{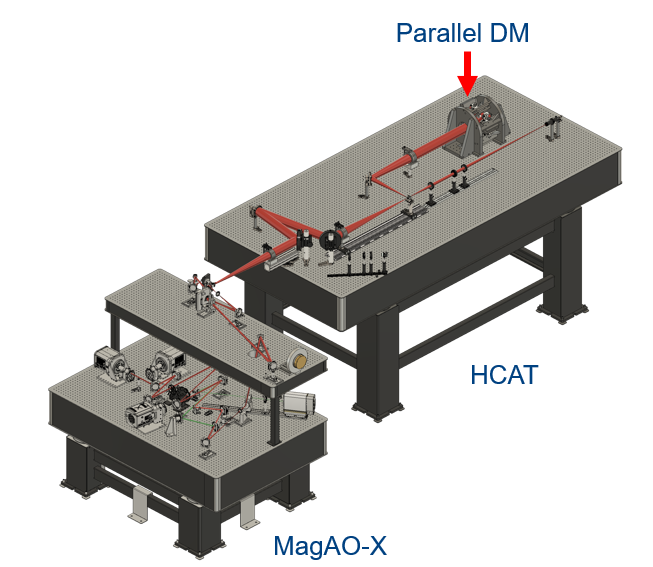}
\end{tabular}
\end{center}
\caption{The CAD rendering shows the HCAT feed to the MagAO-X instrument (Reproduced from Kautz et al. 2023\cite{kautz_gmagao-x_2023}).}
\label{fig:feed} 
    \end{figure}

The lab set up for these experiments utilized both the HCAT bench and the existing ExAO instrument, MagAO-X \edited{(Fig.~\ref{fig:schematic})}. In one room, light is fed from the HCAT table through a hole in a wall with an optical window into the upper tier of the MagAO-X bench \edited{(Fig.~\ref{fig:feed})}.

\subsection{MagAO-X Background}
\label{sec:magaox}
A current extreme adaptive optics (ExAO) system, known as the Extreme Magellan Adaptive Optics system (MagAO-X), was designed for and operates at the 6.5 meter Magellan Clay Telescope at the Las Campanas Observatory in Chile\cite{males_magao-x_2018, close_optical_2018, males_magao-x_2022}. MagAO-X is comprised of two optical benches connected by a periscope relay. The upper bench can be fed directly by the telescope or by an internal source, a super continuum laser (WhiteLase Micro from NKT Photonics) that is fed through a telescope simulator generating an f/11.05 beam which is equal to the focal ratio of the Magellan Clay. On the top bench MagAO-X employs a woofer-tweeter architecture that includes a 97 element ALPAO DM and a Boston Micromachines MEMS 2,040-actuator DM (2k DM) operating up to 3.63kHz (controlled by a PyWFS). On the bottom bench, the lower periscope mirror sends light through a beamsplitter separating the light into the wavefront sensing and science channels. The wavefront sensing path includes a high-speed piezoelectric modulator (PI S-331) and a PyWFS utilizing an EMCCD OCAM\textsuperscript{2}K. The science beam has several filter wheels, including one with a potted HDFS, a low-order wavefront sensor commanding a non-common path corrector DM (NCPC DM), and two science cameras. A recent round of upgrades include a new high-speed low order wavefront sensing camera, a 1,000-actuator non-common path corrector DM (1k NCPC DM), and a new real-time-controller computer\cite{males_magao-x_2022}.

\subsection{HCAT Optical Layout}
\begin{figure} [H]
\begin{center}
\begin{tabular}{c}
\includegraphics[width=0.8\textwidth]{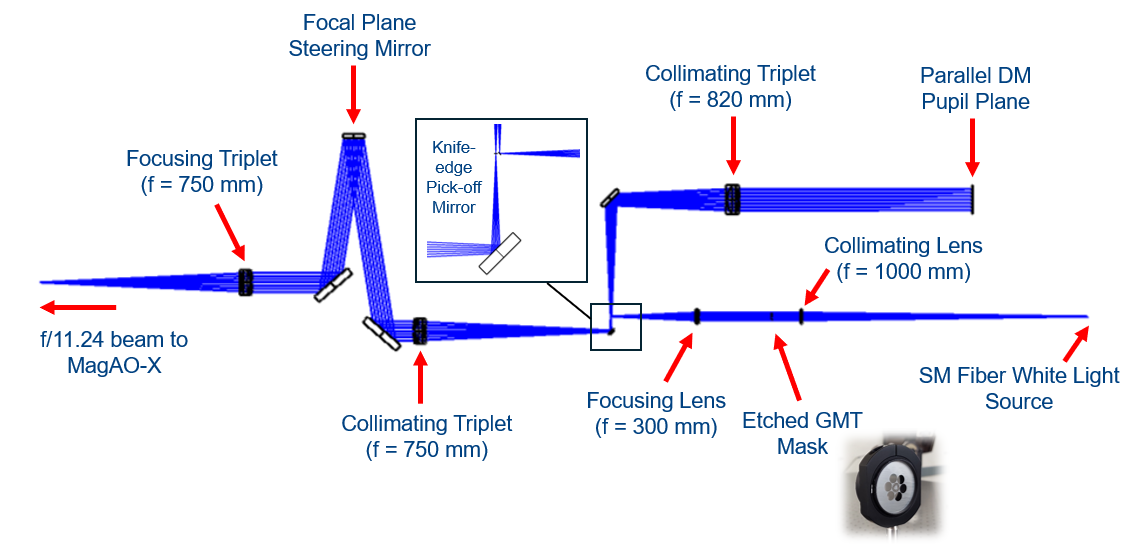}
\end{tabular}
\end{center}
\caption{Zemax rendering of the HCAT optical layout.}
\label{fig:optical} 
    \end{figure}

\begin{figure} [H]
\begin{center}
\begin{tabular}{c} 
\includegraphics[width=0.8\textwidth]{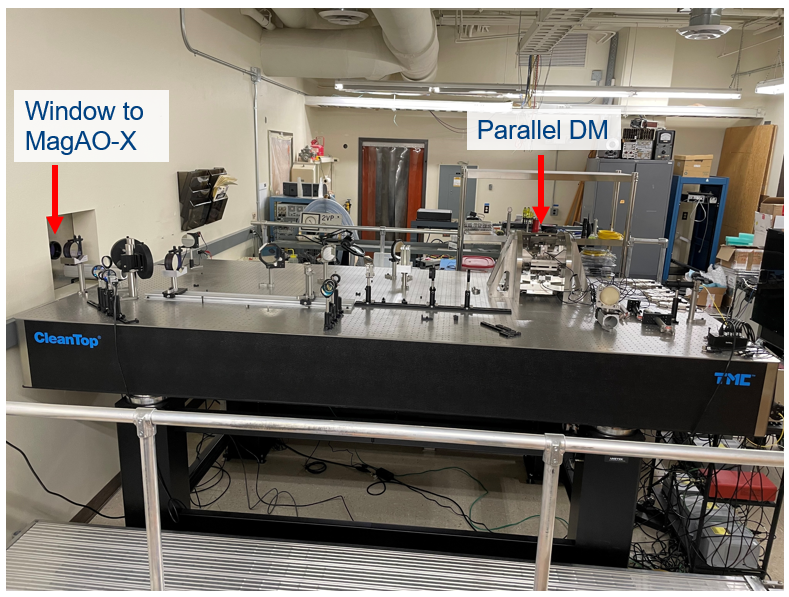}
\end{tabular}
\end{center}
\caption{Built HCAT bench in the Steward Observatory at the University of Arizona.}
\label{fig:built} 
    \end{figure}

\begin{figure} [H]
\begin{center}
\begin{tabular}{c} 
\includegraphics[width=0.8\textwidth]{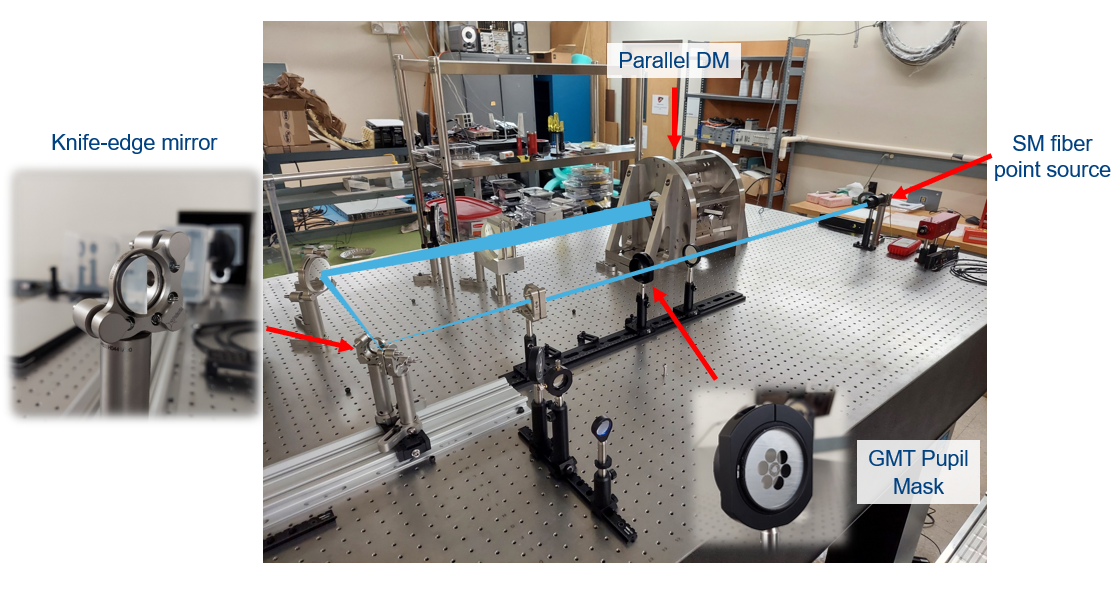}
\end{tabular}
\end{center}
\caption{A $>$1.6 W Thorlabs SLS301 Stabilized Tungsten-Halogen white light source sends light out of a single-mode fiber. The GMT pupil is created \edited{with an etched mask} and sent into the parallel DM and back out.}
\label{fig:hcat_one} 
    \end{figure}

\begin{figure} [H]
\begin{center}
\begin{tabular}{c}
\includegraphics[width=0.9\textwidth]{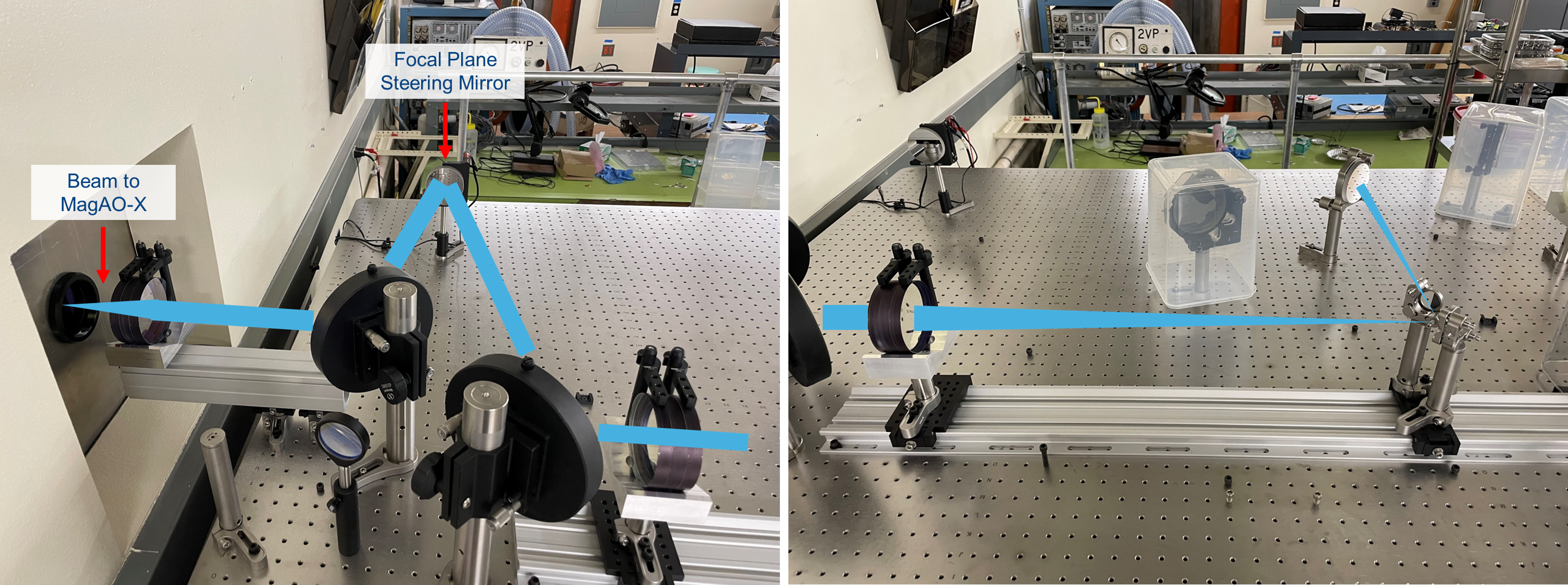}
\end{tabular}
\end{center}
\caption{The focused GMT pupil passes by the knife-edge mirror, onto a fold mirror and into a collimating triplet. The pupil is relayed onto a 3" mirror with piezoelectric tip/tilt control that can act as a focal plane steering mirror. The pupil is refocused and sent through an optical window mounted in a metal plate mounted to the square hole between the HCAT and MagAO-X labs.}
\label{fig:hcat_two} 
    \end{figure}

Figure~\ref{fig:optical} displays the optical layout of the HCAT bench\cite{hedglen_lab_2022}. A GMT pupil is created with an etched mask and that pupil is sent into the ``parallel DM" (see Sec.~\ref{sec:parallel_dm}) and back out in double pass. The coherently recombined beam travels on through the optical system, through an optical window mounted inside a hole in a wall between the HCAT and MagAO-X labs and into the MagAO-X instrument. \edited{~\Crefrange{fig:built}{fig:hcat_two}} show the optomechanical layout of the HCAT bench.

White light is free space coupled into a single-mode fiber (SMF) from a $>$1.6W Thorlabs SLS301 Stabilized Tungsten-Halogen source. The SMF light is collimated then sent through an etched mask of the GMT pupil to simulate the telescope. The light is then focused down to a knife-edge ``D mirror" and sent to a custom collimating triplet \edited{(Fig.~\ref{fig:hcat_one})}. That light is incident on a ``hexpyramid" \edited{within the parallel DM, then} travels through the parallel DM structure and back out in double pass (see Section~\ref{sec:parallel_dm}).

When the light is returning from the parallel DM, it misses the edge of the knife-edge mirror and reflects off of a fold mirror redirecting the light into another collimating triplet \edited{(Fig.~\ref{fig:hcat_two})}.

The light is collimated and a pupil is formed and reflected off of an actuated focal plane steering mirror, then sent through a final f/11.24 focusing triplet through an optical window and into the MagAO-X instrument. Due to pressure differentials between the HCAT and MagAO-X laboratories, this focal plane steering mirror was necessary for maintaining the alignment between the tables (Fig.~\ref{fig:hcat_two}). An f/11.24 beam is sent through an optical window into the MagAO-X instrument (see Sec.~\ref{sec:magaox}). This f-number, slower than the f/11.05 that MagAO-X was designed for, was chosen to slightly undersize the GMT pupil onto the MagAO-X tweeter DM.

\subsection{Parallel DM}
\label{sec:parallel_dm}
\begin{figure}[H]
\centering
\begin{tabular}{c}
\includegraphics[width=0.9\textwidth]{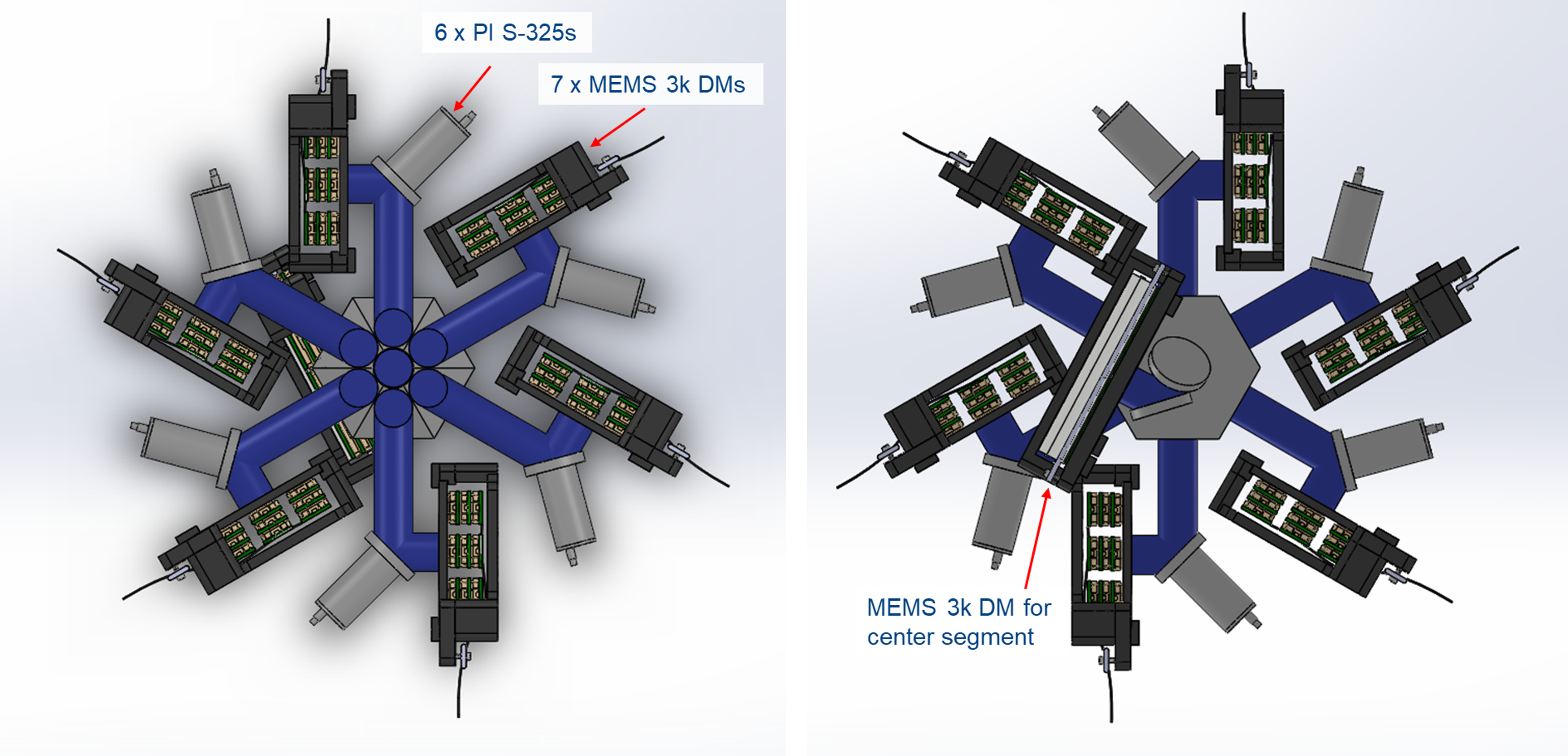}
\end{tabular}
\caption{CAD model of parallel DM concept. \edited{In the front view image on the left, the GMT light} is incident on a six-sided reflective pyramid with a hole in the center, the ``hexpyramid". Each segment is sent outward in a different wavefront control line offset by 60 degrees. Each arm contains a piezoelectic PTT controller and 3k MEMS DM. Due to the double-pass nature of the system, each arm has $\pm$42 $\mu$m of piston OPD stroke. \edited{In the back view image on the right, the} central segment passes through the central hole, onto two crossed fold mirrors and onto a 3k MEMS DM.}
\label{fig:parallel_dm_base}
    \end{figure}

HCAT was developed primarily as a phasing testbed for the Giant Magellean Telescope, to validate the PyWFS + HDFS phasing system. The adaptive secondaries will be the primary phasing control at the GMT and in their absence, a physical system was required to simulate phasing control. Additionally, HCAT is the testbed for experimenting with novel technologies that will be used on the up-and-coming extreme AO instrument for the GMT, GMagAO-X. GMagAO-X is a next-generation instrument that builds upon the heritage of the existing MagAO-X instrument. In order to achieve the same level of wavefront control as the MagAO-X instrument currently does at the Magellan Clay 6.5 m telescope, the density of actuators on the tweeter DM will need to be scaled from the 6.5 m pupil, to the 25.4 m GMT pupil. This scaling means 21,000 actuators are needed to reach an ExAO acceptable wavefront error of $<$90 nm RMS WFE\cite{close_optical_2022}. This can be achieved with seven 3,000 actuator DMs working in parallel, thus the nomenclature ``parallel DM". Within the parallel DM framework, the GMT pupil will be split up by a reflective six-sided pyramid with a central hole, the ``hexpyramid". Each GMT segment will be incident on its own flat fold mirror mounted on a piezoelectric piston/tip/tilt (PTT) controller (Physik Instrumente S-325), then onto its own commercial 3,000 actuator Boston Micromachines (BMC) \edited{MEMS} deformable mirror (DM) that will be employed for phasing and extreme wavefront control \edited{(Fig.~\ref{fig:parallel_dm_base})}. Due to the double pass nature of the system, each arm has $\pm$42 $\mu$m of optical path difference (OPD) dynamic range. See Close et al. 2022\cite{close_optical_2022} for more details about the parallel DM.

\subsubsection{Hexpyramid}
\begin{figure} [H]
\begin{center}
\begin{tabular}{c} 
\includegraphics[height=4cm]{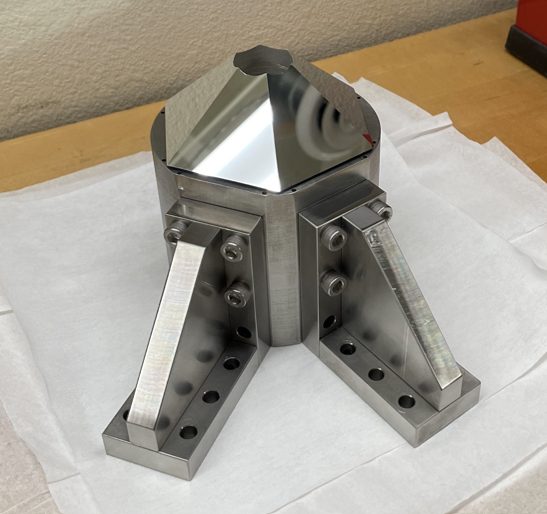}
\end{tabular}
\end{center}
\caption{Manufactured hexpyramid mounted on invar structure to be placed within the parallel DM structure.}
\label{fig:hexpyramid} 
    \end{figure}

In order to create the seven distinct wavefront control lines for each GMT segment, the GMT pupil will be incident upon a reflective six-sided pyramid with a hole in the center called the ``hexpyramid" (Fig.~\ref{fig:hexpyramid}). The six outer segments, will be sent outward in six different directions towards a corresponding piezoelectric PTT actuator and MEMS DM while the central segment passes through the center to a MEMS DM. The center segment will not have a designated piezoelectric actuator as it will be used as the reference segment for correcting differential piston.

Various optical arrangements of the DMs were considered but ultimately a crossed-fold design was selected due to the higher Strehl images produced\cite{hedglen_lab_2022}.

\begin{figure}[H]
\centering
\begin{tabular}{c}
\includegraphics[width=0.9\textwidth]{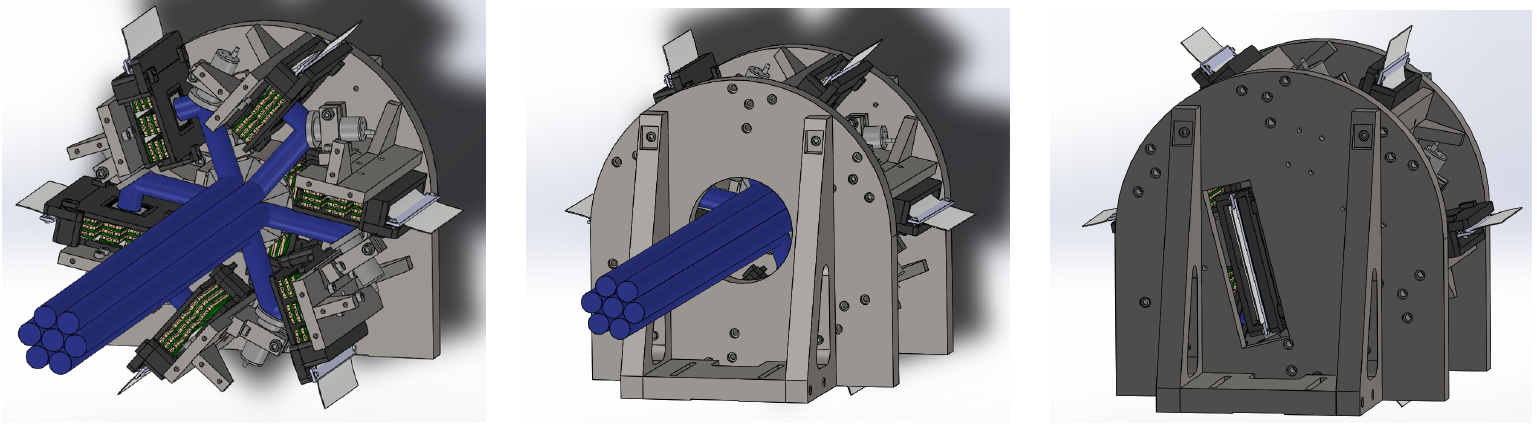}
\end{tabular}
\caption{CAD renditions of parallel DM to be used on GMagAO-X. The left image shows the structure without its front plate to clarify the ray path. The middle image shows the structure with its front plate on. The right image shows the back of the structure.}
\label{fig:parallel_dm_CAD_closed}
    \end{figure}

Figure~\ref{fig:parallel_dm_CAD_closed} shows the CAD models of the front and back of the structure around the parallel DM. There is a hole in the back of the structure for the final DM corresponding to the central segment.

\begin{figure}[H]
\centering
\begin{tabular}{c}
\includegraphics[height=8cm]{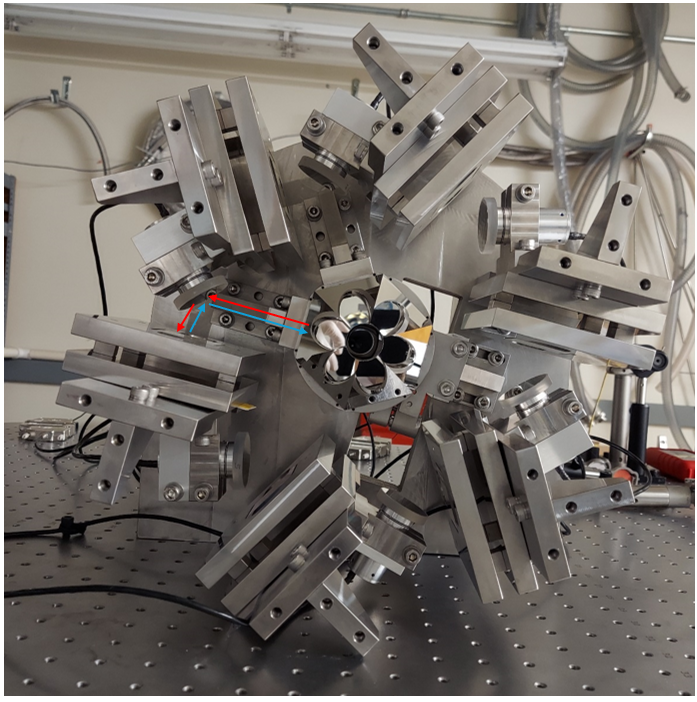} 
\end{tabular}
\caption{This is the as-built parallel DM on HCAT bench with the six piezoelectric PTTs, \edited{six ``mock DMs", and two fold flats plus a kinematically mounted flat for the central segment (behind hexpyramid).} The ray path from hexpyramid to PTT controller to mock DM and back is depicted in red and blue arrows respectively.}
\label{fig:parallel_dm_ray_path}
    \end{figure}

As the HCAT project does not currently have funding for \edited{seven} 3k MEMS DMs, the current ``parallel DM" on the HCAT bench includes \edited{six} PI S-325 piezoelectric piston/tip/tilt actuators and \edited{seven} flat mirrors currently in place of the MEMS devices, known as ``mock DMs". This is sufficient for the in-lab phasing experiments where the parallel DM \edited{PTTs act as the surrogates for the GMT's ASMs. The central reference segment is stationary and does not require a PTT. Figure~\ref{fig:parallel_dm_ray_path} shows how light would travel through the built parallel DM system in double pass.}

\subsubsection{Bypass Mode}
\begin{figure} [H]
\begin{center}
\begin{tabular}{c} 
\includegraphics[height=6cm]{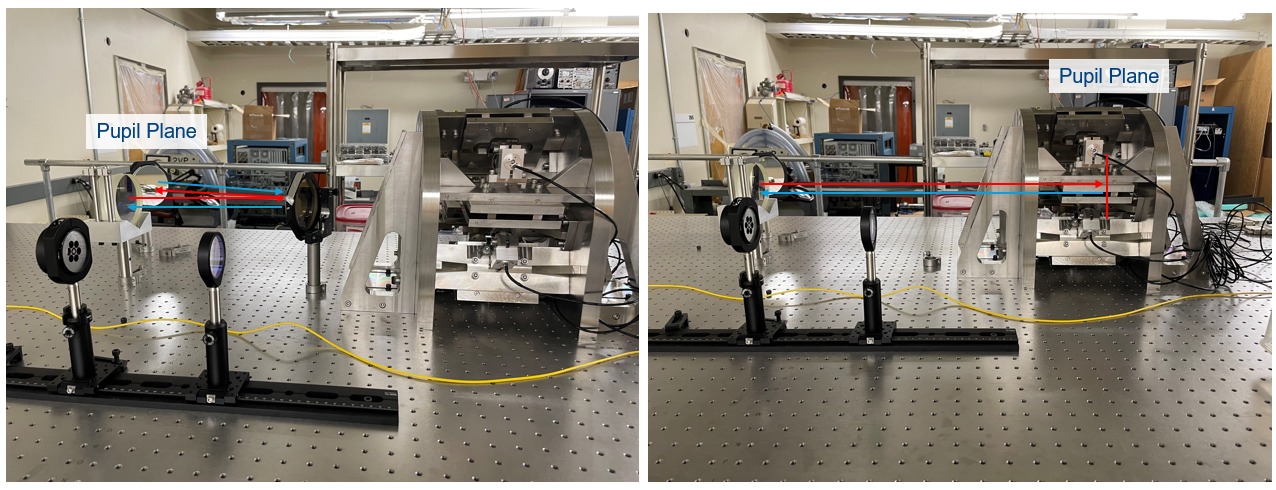}
\end{tabular}
\end{center}
\caption{In ``Bypass Mode", on the left, two fold mirrors are placed ahead of the parallel DM so the phased GMT pupil is relayed in and out, in place of an unphased pupil created by the parallel DM. ``Parallel DM Mode", on the right, shows the standard ray path of the GMT pupil entering and exiting the parallel DM.}
\label{fig:bypass} 
    \end{figure}

HCAT can operate in two modes: bypass mode and parallel DM mode\cite{hedglen_development_2023}. In bypass mode, two fold mirrors are placed ahead of the parallel DM to ``bypass" it and relay a perfectly phased GMT pupil (ie no splitting and recombining) to the rest of the testbed. Bypass mode is used for alignment to MagAO-X and will be utilized for creating reference point spread functions (PSFs). In parallel DM mode, the fold mirrors are removed, the \edited{individual segment pupil planes are on the mock DMs}, and HCAT utilizes the PTTs for phasing experiments. \edited{The two modes are shown in Figure~\ref{fig:bypass}.}

\section{In-lab Phasing Experiments}
\label{sec: in_lab}
\subsection{Calibration Set-up}
\begin{figure} [H]
\begin{center}
\begin{tabular}{c} 
\includegraphics[width=0.8\textwidth]{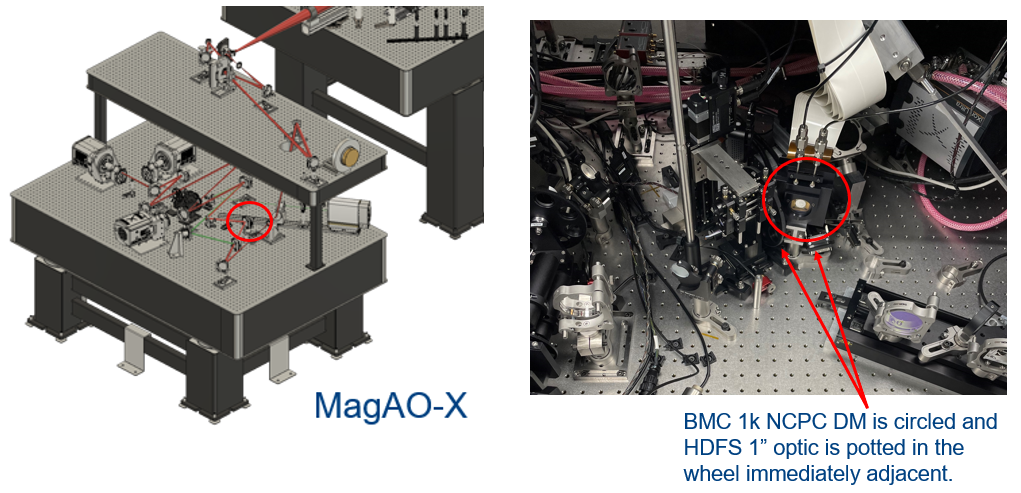}
\end{tabular}
\end{center}
\caption{The 1,000 actuator MEMs non-common path corrector (NCPC) DM is placed in the science beam on the lower bench of MagAO-X. Immediately following it is the pupil filter wheel housing the HDFS.}
\label{fig:ncpc} 
    \end{figure}

The HDFS was calibrated using a cross-correlation template method \cite{haffert_phasing_2022}. In this method, we step through a ramp of piston values for each pair of segments. The images for each differential piston value are saved in a library. To create the reference library of piston errors, HCAT was put into bypass mode and a perfectly phased GMT pupil is fed into MagAO-X. We segmented the NCPC DM on MagAO-X \edited{(Fig.~\ref{fig:ncpc})} to match the seven GMT segments incident on the DM and swept through the full \edited{$\sim$2.5} $\mu$m \edited{optical} range of the NCPC DM in steps of roughly 10 nm. This constructs a library of reference images that can be cross-correlated with HDFS images produced by an unphased pupil\cite{haffert_phasing_2022}.

The differential piston reconstruction starts by masking a particular pair of fringes (the $m = -1$ and $m = +1$ orders). These are then normalized by subtracting the mean and dividing by the standard deviation within the mask. The correlation signal is calculated by taking the inner product between the normalized fringes and the corresponding fringe reference library images. This results in a correlation as a function of the calibrated differential piston. The most likely differential piston value is found by determining the peak of the correlation function. The peak is found by fitting a 2nd-order polynomial around the peak pixel and then deriving the peak position of the 2nd-order polynomial. This approach allows us to get a better precision than the original calibration precision of 10 nm.

\subsection{Phasing with the NCPC DM}
\begin{figure} [H]
\begin{center}
\begin{tabular}{c} 
\includegraphics[height=8cm]{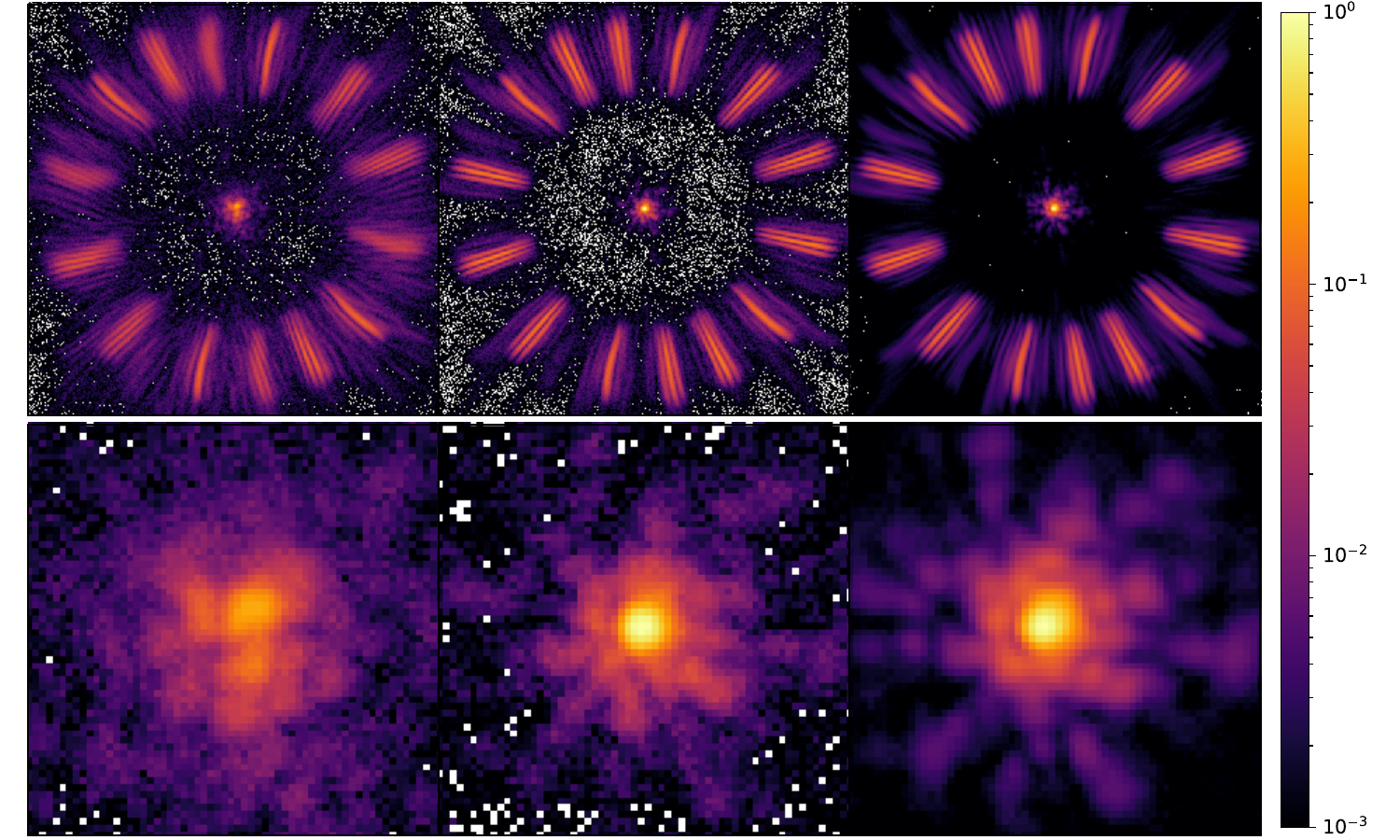}
\end{tabular}
\end{center}
\caption{The left column shows the unphased HDFS image and PSF image when a piston scramble is input onto the NCPC DM. The center column shows the HDFS image and PSF image when the NCPC DM had phased itself. The right column shows the reference PSF used to calculate the relative Strehl of the phased image. The lower row log-stretch PSFs are from zooming in on the PSF in the center of the broadband HDFS images (600-950 nm).}
\label{fig:unphased_phased_ncpc} 
    \end{figure}

We experimentally demonstrate closed-loop piston control of the segmented NCPC DM with our calibration \edited{(Fig.~\ref{fig:unphased_phased_ncpc})}. The central segment was held at a constant piston and various pistons up to \edited{$\sim$2.5} $\mu$m peak-to-valley WFE were input on the other segments to be sensed by the HDFS and corrected. The PyWFS + MagAO-X tweeter DM were used to correct bench seeing.

\begin{figure} [H]
\begin{center}
\begin{tabular}{c}
\includegraphics[width=0.7\textwidth]{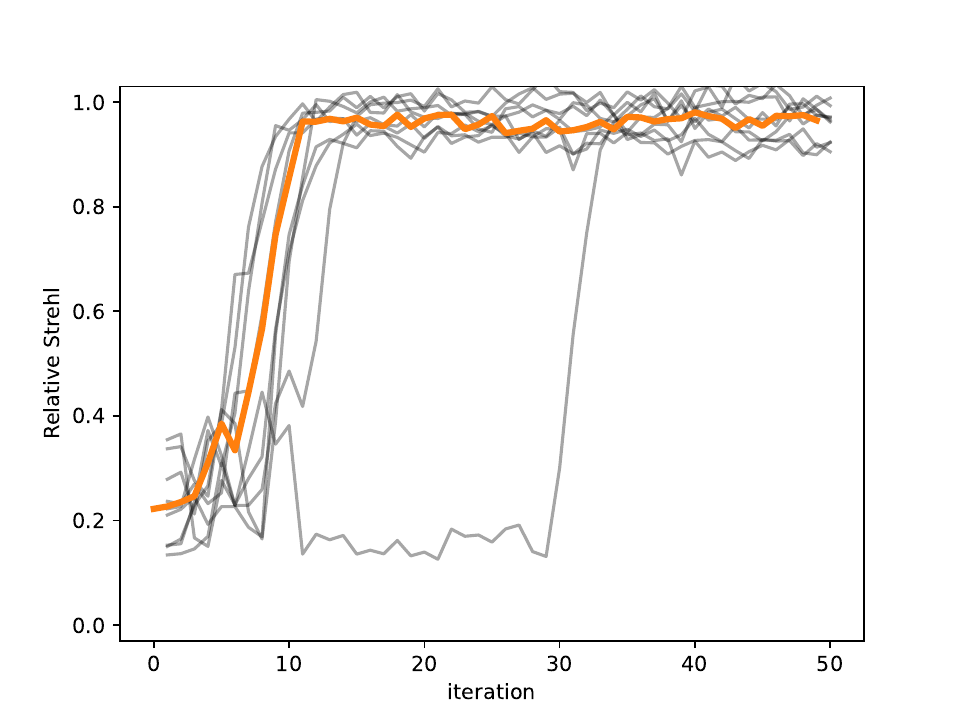}
\end{tabular}
\end{center}
\caption{The ``relative Strehl" is defined as the ratio between the encircled energy within 1$\lambda$/D of the broadband PSF at the end of the phasing experiments to that of the reference broadband PSF.}
\label{fig:rel_strehl} 
\end{figure}

We took a reference ``phased" PSF before beginning the closed loop experiments. This allowed us to take a ``relative Strehl" measurement. First we created a photometric mask 6 pixels in diameter (roughly 1$\lambda$/D) centered on the unaberrated PSF and measured the encircled energy. We measured the relative Strehl as the ratio between the encircled energy at the end of the phasing experiments and the encircled energy of the reference PSF without phasing errors. We did a relative Strehl measurement because we were only controlling differential piston during the experiment. \edited{Small drifts in the system introduced low-order non-common path aberrations (NCPA) that were also controlled and removed by the HDFS because certain low-order modes can be partially compensated by differential piston.} The median relative strehl was 96.8$\% \pm3\%$ (Fig.~\ref{fig:rel_strehl}). We estimate the residual piston error to be $22_{-17}^{+9}$ nm RMS at $\lambda$ = 760 nm based on the Maréchal approximation. This falls well within our goal of correcting residual piston to within $\pm\lambda$/2. \edited{}

\subsection{Phasing with HCAT and the Parallel DM}
\begin{figure} [H]
\begin{center}
\begin{tabular}{c} 
\includegraphics[height=8cm]{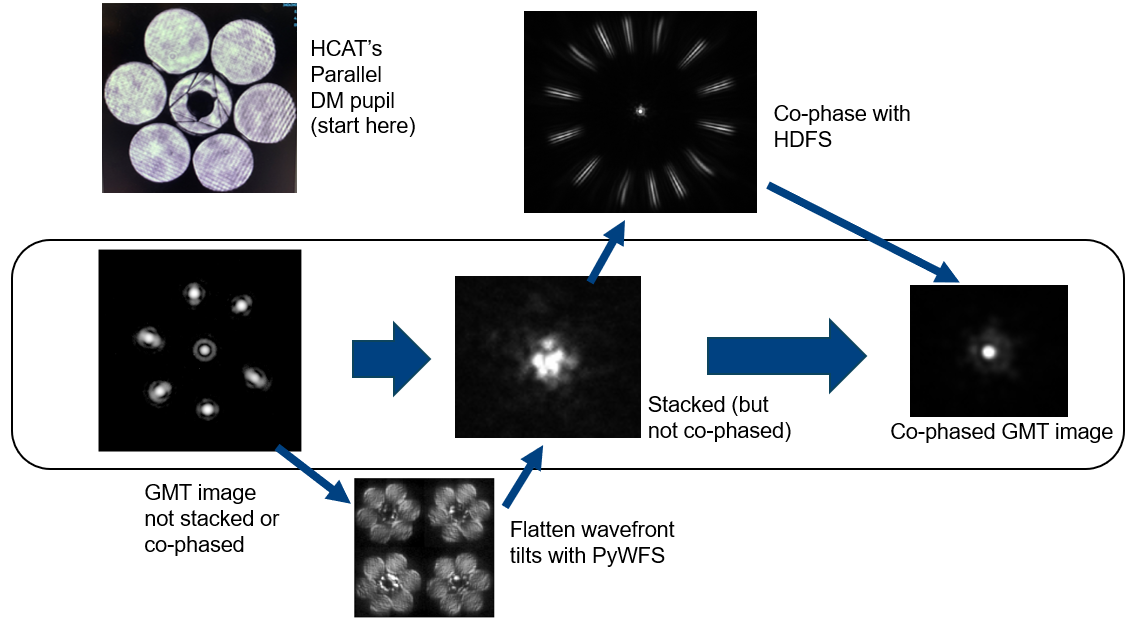}
\end{tabular}
\end{center}
\caption{Like the unphased GMT, the parallel DM begins in a state of seven distinct PSFs. The MagAO-X PyWFS is used to sense tip/tilt so the PTTs can stack the PSFs onto one another into a singular incoherent (ie unphased) PSF. The HDFS senses differential piston so the PTTs can then perform the phasing to achieve a final coherent PSF.}
\label{fig:phasing} 
    \end{figure}

When the parallel DM is turned on it is in an unphased state. Although each PI S-325 is operating in closed loop about its commanded position, after power cycling the previous absolute calibration is changed. The system begins with seven distinct PSFs in the focal plane. These need to be stacked into one singular PSF. We utilized the MagAO-X PyWFS to sense these tip/tilt errors and use the tip/tilt control of the piezos to perform the stacking. The PyWFS also monitors bench seeing and sends controls back to the tweeter DM. The HDFS senses the piston errors, we then utilize the reference library to cross-correlate the images and determine the differential piston, then send the corrective piston command back to the parallel DM. \edited{In this experiment, the pyramid is not used for piston sensing. Figure~\ref{fig:phasing} shows the process of going from an unphased GMT pupil to a final phased GMT PSF using HCAT's parallel DM and MagAO-X.}

\begin{figure} [H]
\begin{center}
\begin{tabular}{c}
\includegraphics[width=\textwidth]{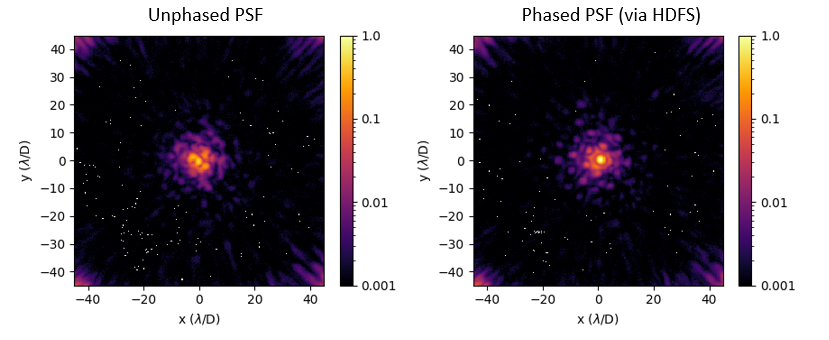}
\end{tabular}
\end{center}
\caption{GMT PSF before and after phasing the parallel DM using the HDFS. Images are taken in MagAO-X’s z' filter: 908 nm (130 nm bandwidth).}
\label{fig:pdm_result} 
    \end{figure}

Figure~\ref{fig:pdm_result} shows the PSF before and after phasing the parallel DM. Using a model-based Strehl estimate, we measure an absolute Strehl of approximately 35$\%$ on the z' 908 nm PSF. Further close-loop control experiments in the presence of turbulence will be required to prove the validity of the HDFS + parallel DM combination for phasing control on GMagAO-X.

\subsubsection{Collaboration with GMT NGWS-P Team}
\begin{figure} [H]
\begin{center}
\begin{tabular}{c} 
\includegraphics[width=\textwidth]{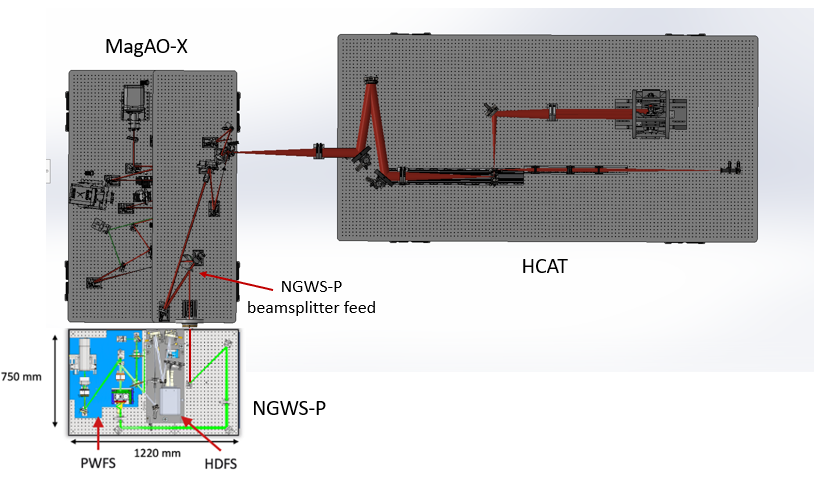}
\end{tabular}
\end{center}
\caption{In order to validate PyWFS + HDFS hardware and software for the GMT's Natural Guide star Wavefront Sensing Prototype (NGWS-P), the system was fed by the HCAT GMT simulator in conjunction with the MagAO-X instrument. An f/57 beam reflected off of a beamsplitter exits the MagAO-X eyepiece and is periscoped down into the NGWS-P beam path (periscope not shown).}
\label{fig:ngwsp} 
    \end{figure}

\edited{Another aspect} of the HCAT project is to feed light from HCAT, through MagAO-X, to the Natural Guide star Wavefront Sensor Prototype (NGWS-P) (Fig.~\ref{fig:ngwsp})\cite{plantet_ngws-p_2022}. NGWS-P is the prototype phasing testbed for the official GMT NGWS system. It was built in a collaboration between the Giant Magellan Telescope Organization and the Osservatorio Astrofisico di Arcetri in Florence, Italy.

The main goal of the collaboration is for GMTO to verify its internal PyWFS + HDFS architecture and control algorithms. There have been three NGWS-P runs with HCAT at the University of Arizona to date. \edited{The NGWS-P bench has its own PyWFS and HDFS for differential piston sensing.} In the experiments, the MagAO-X 2k DM and HCAT's six PTTs are used in place of the GMT adaptive secondaries for wavefront control and phasing control respectively.

\edited{Using the parallel DM's six PTTs in combination with the HDFS and PyWFS controlling differential piston modes and bench seeing, the NGWS-P achieved a Strehl of 73$\%$ at 850 nm. For more in depth discussion of these experiments see Plantet et al. and Quirós-Pacheco et al.\cite{plantet_integration_2024, quiros-pacheco_giant_2024}} 

\section{On-sky Phasing Experiments with the HDFS}
\label{sec: on_sky}
\subsection{Phasing with  MagAO-X and the NCPC DM in On-sky Turbulence with the HDFS}
\begin{figure} [H]
\begin{center}
\begin{tabular}{c} 
\includegraphics[height=3.6cm]{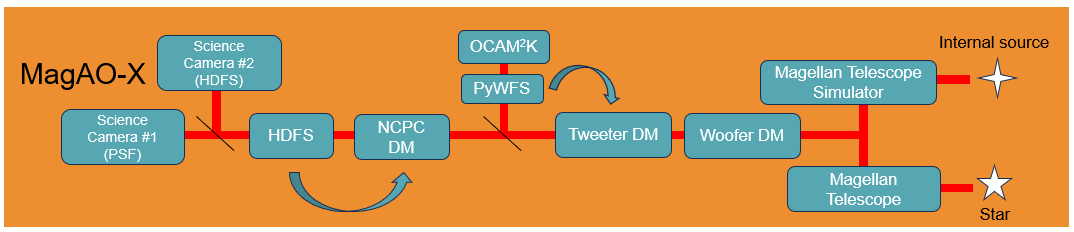}
\end{tabular}
\end{center}
\caption{For daytime calibrations, we used the MagAO-X internal white light source with the Magellan pupil mask in our internal telescope simulator. For on-sky experiments, MagAO-X is fed starlight directly by the 6.5 m Magellan Clay Telescope. The NCPC DM is used to create segment piston to be sensed by the HDFS.}
\label{fig:schematic_onsky} 
    \end{figure}

We wanted to demonstrate HDFS phasing in actual seeing conditions with real residual AO WFE. When MagAO-X was brought to the Magellan Clay for its March 2024A run, we recreated the NCPC closed-loop phasing test. \edited{At the telescope, feeding in HCAT's bypass mode GMT pupil is not possible, we have to use starlight or the MagAO-X internal source (Fig.~\ref{fig:schematic_onsky}).} Before going on-sky, we used the MagAO-X internal source, MagAO-X internal telescope simulator (of the Magellan Clay), and the NCPC DM to create a six segment pupil to perform phasing tests. As it was mostly blocked by the Magellan Clay's central obscuration, the seventh segment is not included.

\begin{figure} [H]
\begin{center}
\begin{tabular}{c} 
\includegraphics[width=0.8\textwidth]
{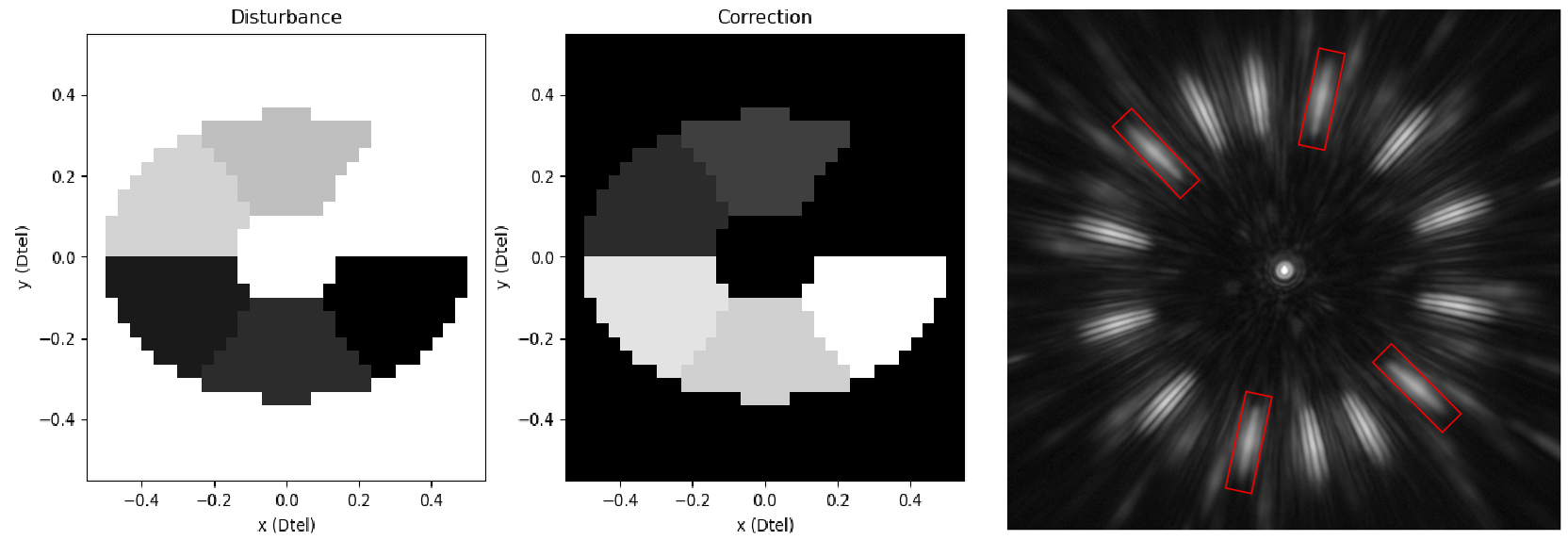}
\end{tabular}
\end{center}
\caption{(left) The pattern imprinted on the NCPC DM to segment the Magellan pupil. \edited{The projected differential piston pattern is asymmetric because of the 45 degree incidence angle.} (right) Reference HDFS image taken with MagAO-X internal white light source. The four barber poles without interference are enclosed by red rectangles.}
\label{fig:reference_hdfs_chile} 
    \end{figure}

Since a six segment pupil was used instead of the seven segment GMT pupil, the focal plane of the HDFS only has 10 fringes. The four extra barber poles are present but since there is no seventh central segment, there is no interference present \edited{(Fig.~\ref{fig:reference_hdfs_chile}).}

\begin{figure} [H]
\begin{center}
\begin{tabular}{c} 
\includegraphics[width=\textwidth]{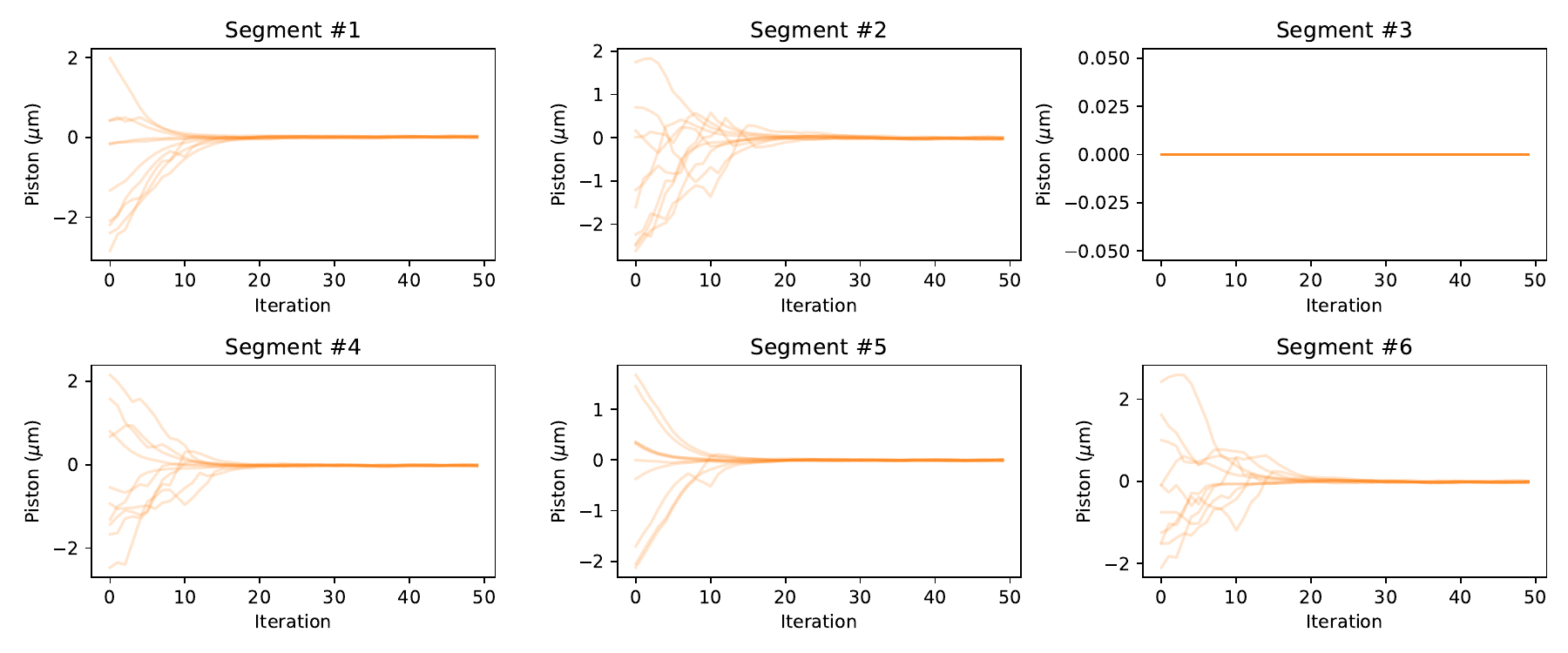}
\end{tabular}
\end{center}
\caption{Differential piston was input onto five segments, sensed by the HDFS, and corrected by the NCPC DM.}
\label{fig:ncpc_inchile} 
    \end{figure}

\begin{figure} [H]
\begin{center}
\begin{tabular}{c} 
\includegraphics[width=\textwidth]{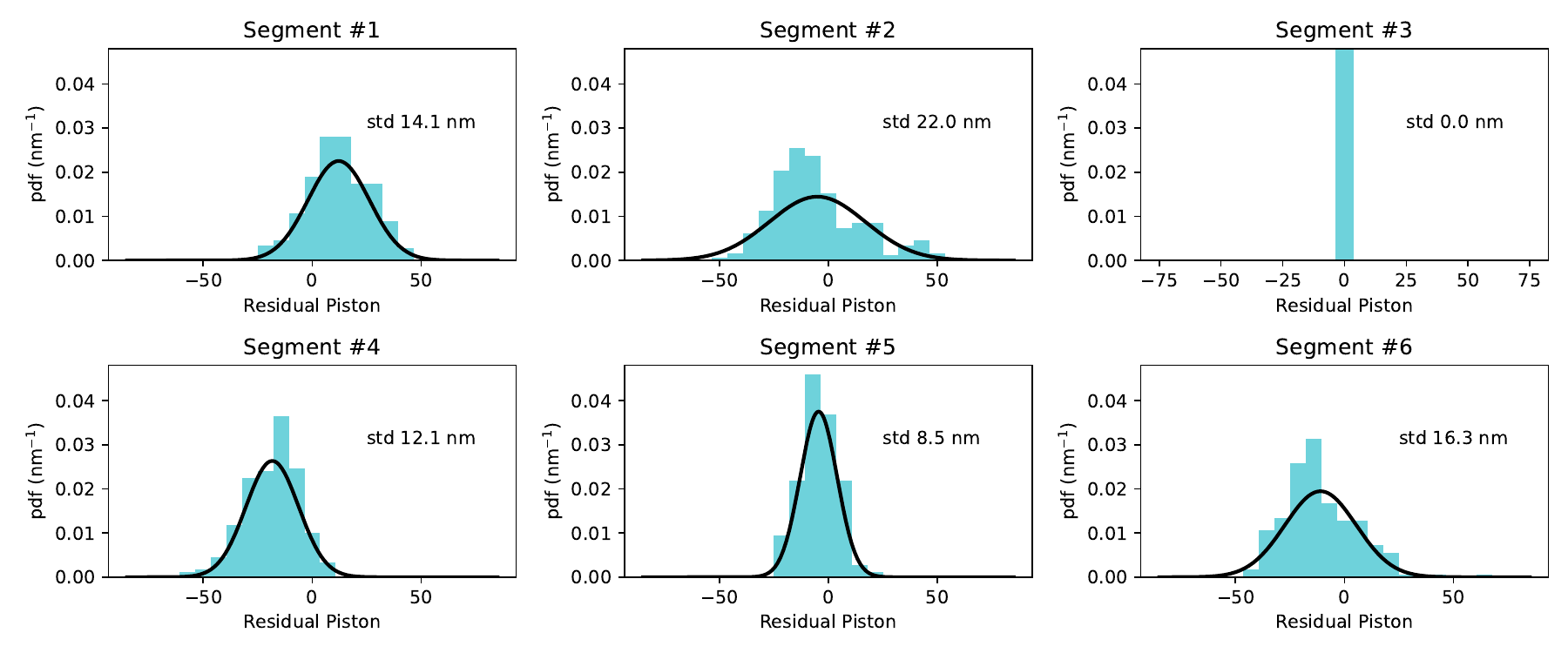}
\end{tabular}
\end{center}
\caption{These histograms show the peak-to-valley residual piston on each segment after iteration 25. A Gaussian distribution matching the standard deviation of each histogram is
plotted for each histogram. The residuals are at the nm level. Only internal ``bench seeing" turbulence was present.}
\label{fig:ncpc_inchile_hist} 
    \end{figure}

Optical systems are not sensitive to absolute phase. Therefore, there is a degeneracy if we apply differential piston to all of the six segments. One of the segments must be held constant as the reference phase segment. For the experiments here, Segment $\#$3 was used as the phase reference segment to create the daytime calibration reference library \edited{(Fig.~\ref{fig:ncpc_inchile} and Fig.~\ref{fig:ncpc_inchile_hist})}.

The experiment for phasing stability was repeated using just the internal source and the NCPC DM.
We were able to control phase to within $\pm\lambda$/11.3 at $\lambda$ = 800 nm in closed loop (Fig.~\ref{fig:ncpc_inchile}). The residual distributions are not centered on zero. We attribute this to slow NCPA drifts between when the HDFS was calibrated and when the closed-loop experiments were done. There was a time difference of several hours. We had to realign the PyWFS in between the calibration and experiment. This can lead to new NCPA errors. The residuals correspond to about 30 to 40 nm rms of low-order wavefront error. \edited{Changes in the alignment of the PyWFS could potentially create this amount of wavefront error}.

Next, we demonstrated on-sky closed-loop control of the segmented Magellan aperture. \edited{The on-sky experiments were to show the control of static phase errors in the presence of partially compensated atmospheric turbulence. The PyWFS was not used for phasing.} The star used was ``a Cen" (HD 125823) which has an I magnitude of 4.7, spectral type B2V. The tweeter loop was operating at 2kHz and \edited{the PyWFS} was modulating at 3$\lambda$/D. The HDFS is effectively a focal plane wavefront sensor. This means that residual atmospheric turbulence could impact the reconstruction. Therefore, we decided to take long exposures (1 to 10 seconds) to average out the residual atmospheric turbulence not corrected by the 2kHz PyWFS AO loop. The long exposure incoherent halo then wouldn't impact the reconstruction. The HDFS fringes were imaged onto the 1024x1024 pixel EMCCD science camera (Fig.~\ref{fig:schematic_onsky}). For the actual observations itself, we used a smaller region of interest of 512x512 pixels that was centered on the 0\textsuperscript{th} order of the HDFS.

\begin{figure} [H]
\begin{center}
\begin{tabular}{c} 
\includegraphics[width=\textwidth]{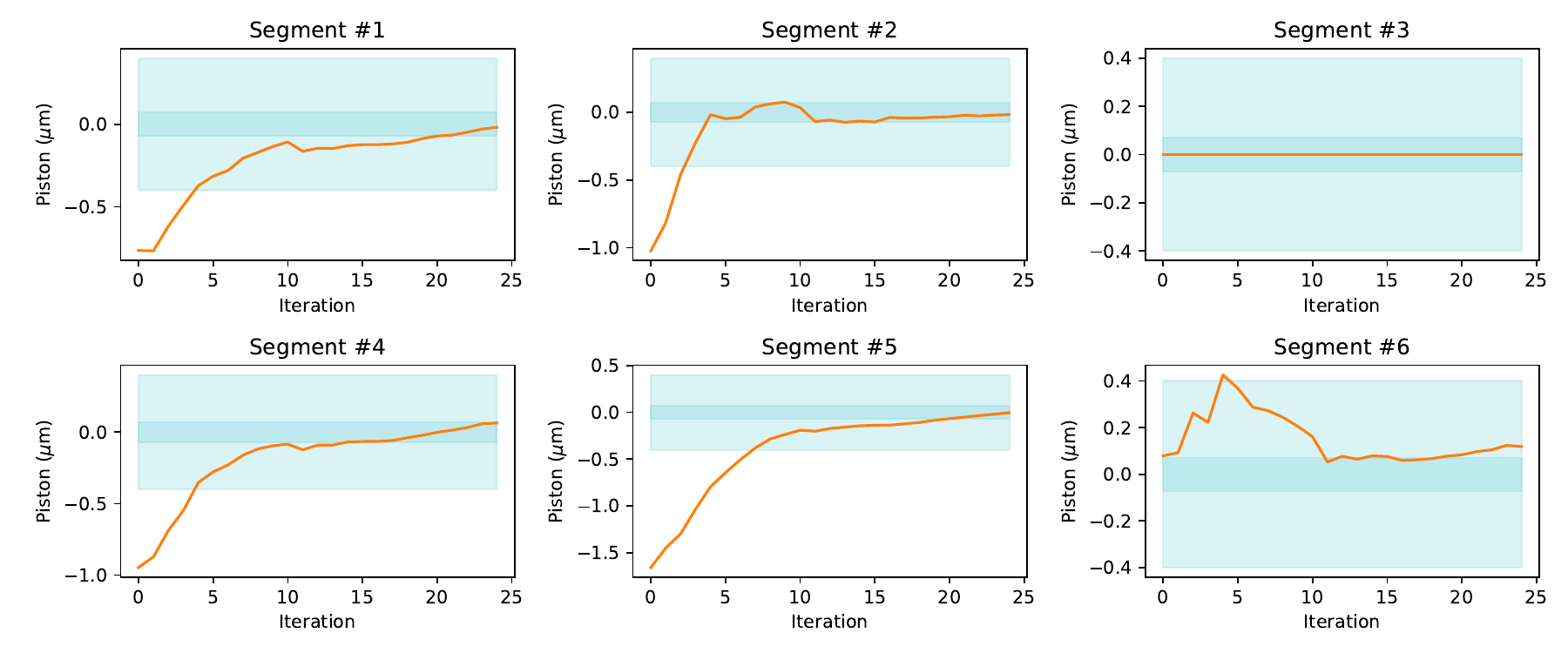}
\end{tabular}
\end{center}
\caption{Trial $\#$1 had 25 iterations and four of the five segments with input piston converged to $\pm\lambda$/11.3 at $\lambda$ = 800 nm. Seeing was 0.6".}
\label{fig:ncpc_inchile_onsky_1} 
    \end{figure}

\begin{figure} [H]
\begin{center}
\begin{tabular}{c} 
\includegraphics[width=\textwidth]{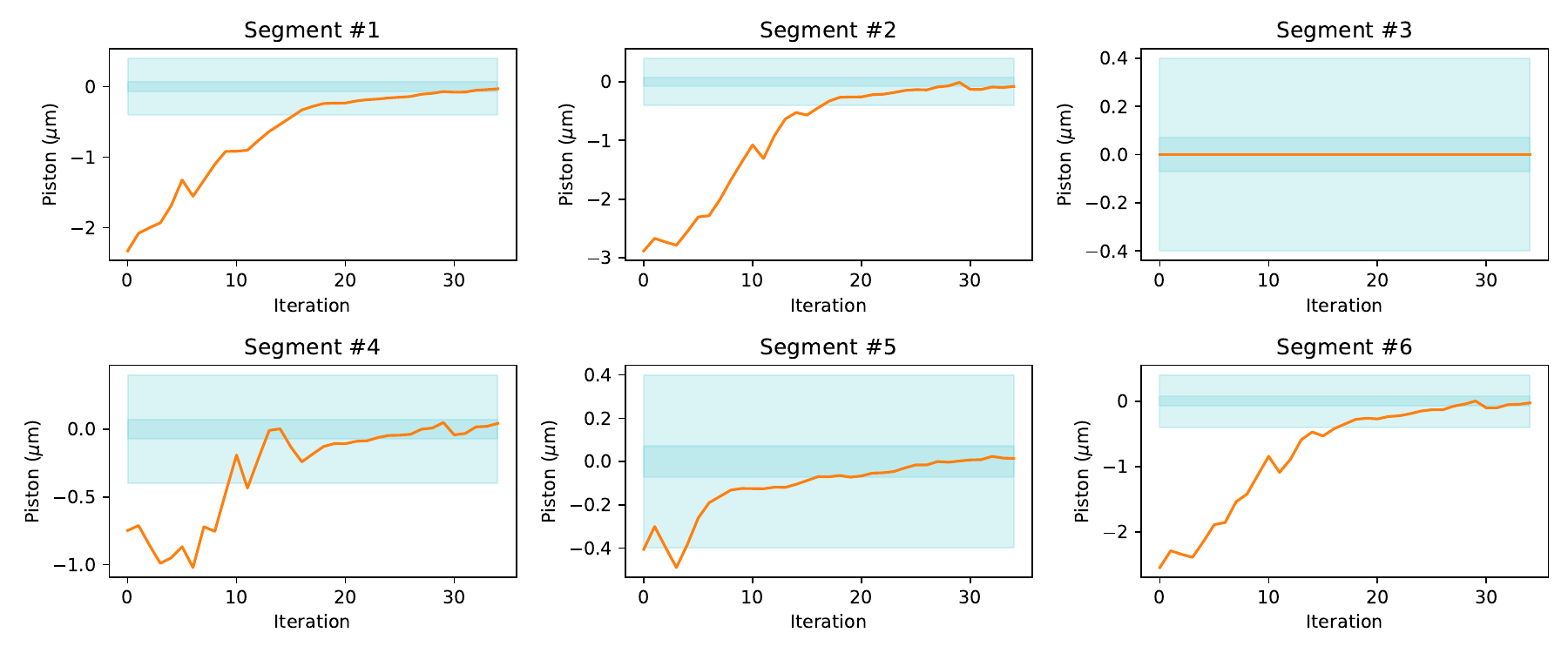}
\end{tabular}
\end{center}
\caption{Trial $\#$2 had 35 iterations and all five segments with input piston converged to $\pm\lambda$/11.3 at $\lambda$ = 800 nm. Seeing was 0.7".}
\label{fig:ncpc_inchile_onsky_2} 
    \end{figure}

\begin{figure} [H]
\begin{center}
\begin{tabular}{c}
\includegraphics[width=\textwidth]{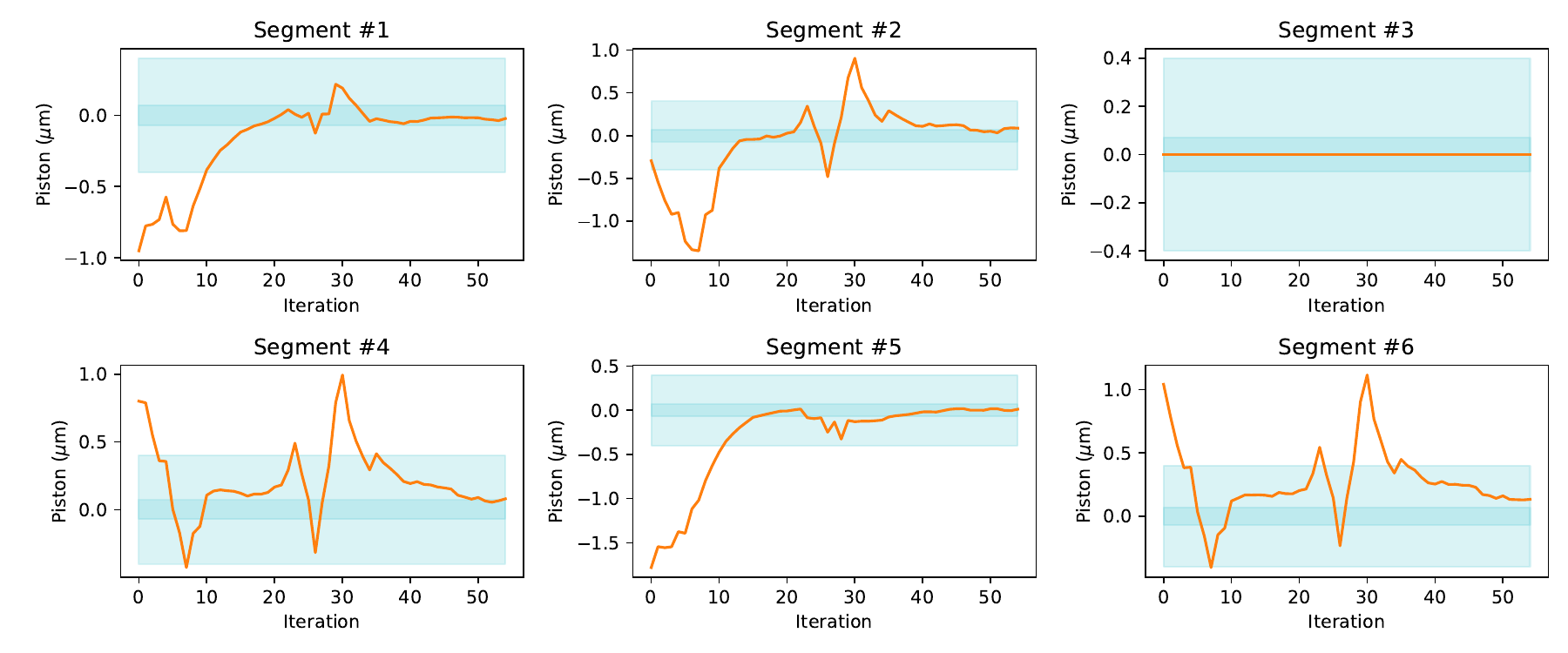}
\end{tabular}
\end{center}
\caption{Trial $\#$3 had 55 iterations and four of the five segments with input piston converged to $\pm\lambda$/11.3 at $\lambda$ = 800 nm. Seeing was 0.89" and jumped to 1.03" by the end of the trial. There is a noticeable burst of seeing around iteration 30.}
\label{fig:ncpc_inchile_onsky_3} 
    \end{figure}

We performed three phasing trials. We input a piston scramble of up to \edited{$\sim$2.5} $\mu$m peak-to-valley WFE by saturating the DM on five of the six segments, holding Segment $\#$3 constant as the reference segment. Trial $\#$1 had 25 iterations with seeing around 0.6". Trial $\#$2 had 35 iterations with seeing around 0.7". The final Trial $\#$3 had 55 iterations with seeing jumping from 0.89" to 1.03" by the end of the trial. \edited{The burst of seeing around iteration 30 created a notable disturbance in the control loop. This caused the differential piston control to be lost momentarily, as shown by a spike of piston error on each segment.} \edited{Seeing was measured from the telescope site's differential image motion monitor (DIMM).} ~\Crefrange{fig:ncpc_inchile_onsky_1}{fig:ncpc_inchile_onsky_3} illustrate the HDFS successfully phasing to $<$141 nm peak-to-valley WFE, or $\sim$50 nm RMS WFE integrated across the whole pupil, (which means $<$$\pm$$\lambda$/11.3 at $\lambda$ = 800 nm from the reference piston segment to each segment) in median to poor seeing conditions. The light blue shaded region indicates the $\pm$$\lambda$/2 WFE goal region ($\lambda$ = 800 nm). The darker blue shaded region indicates the $\pm$$\lambda$/11.3 WFE ($\lambda$ = 800 nm) we were able to successfully phase to.

\begin{figure} [H]
\begin{center}
\begin{tabular}{c}
\includegraphics[width=\textwidth]{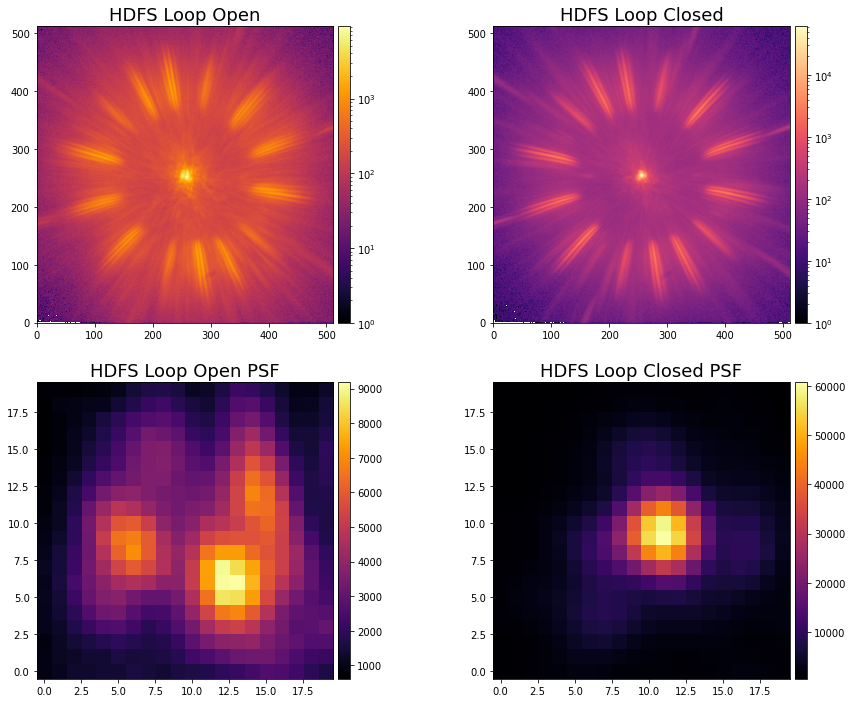}
\end{tabular}
\end{center}
\caption{The left images show the unphased HDFS in open-loop and zoomed-in linear stretch PSF from Trial $\#$2. The right images show the corresponding phased HDFS in closed-loop and zoomed-in linear stretch PSF. To be clear, all images are taken with the high-order AO loop closed, just the feedback loop from the HDFS piston sensor is toggled off/on. We calibrated the spectral dispersion of the HDFS with a set of narrowband filters and found the extent to be 530 nm - 1070 nm.}
\label{fig:on_sky_hdfs} 
    \end{figure}

\edited{Figure~\ref{fig:on_sky_hdfs} shows images of the HDFS focal plane in open and closed loop on-sky. The PSFs in the center of the image are enlarged to show the effect of correcting differential piston errors.}

\section{CONCLUSION}
All three Giant Segmented Mirror Telescopes will have a new challenge of phasing the differential segment piston. \edited{Pupil fragmentation due to secondary struts on E-ELT and TMT will induce petal modes via the low wind effect, and/or the isolated island effect of their wavefront sensors\cite{salama_keck_2024, schwartz_sensing_2017, bertrou-cantou_confusion_2022}}. Specifically on the GMT, the gaps between segments are larger than an atmospheric coherence length, so standard wavefront sensors will not be usable for sensing differential segment piston \edited{petal modes}. The GMT has chosen a two-channel phasing system comprised of a holographic dispersed fringe sensor (HDFS) and a pyramid wavefront sensor (PyWFS) to do the coarse and fine phasing respectively. \edited{Theoretically, an HDFS could be redesigned for the E-ELT and TMT pupils to sense differential piston between the large petals created by the secondary struts.} We present the initial lab and on-sky phasing demonstrations using the HDFS and the High Contrast Adaptive optics phasing Testbed and the ExAO instrument, MagAO-X. The HDFS successfully phased to $<$141 nm peak-to-valley WFE, or $\sim$50 nm RMS WFE integrated across the whole pupil, (which means $<$$\pm$$\lambda$/11.3 at $\lambda$ = 800 nm from the reference piston segment to each segment) in median to poor seeing conditions. This would be a sufficient correction to hand off to a PyWFS to complete the fine phasing of the segments and is a very promising result for GMT NGAO. The ``parallel DM" structure of multiple piezoelectric controllers on the HCAT table allows us to further simulate phasing at the GMT. Next steps include demonstrating robust closed loop control of differential piston with the parallel DM using HDFS + PyWFS feedback in turbulence. This will allow us to furthur probe GMT's NGAO strategy and verify the parallel DM itself as a phasing controller for the up-coming visible/NIR ExAO GMT instrument, GMagAO-X.

\section{Code and Data Availability Statement}
Laboratory data was obtained from the MagAO-X instrument in-lab at the University of Arizona and at the Las Campanas Observatory. Data is available from the authors upon request.

\section{Disclosures}
The authors declare there are no financial interests, commercial affiliations, or other potential conflicts of interest that have influenced the objectivity of this research or the writing of this paper.
 
\acknowledgments 
 
The HCAT testbed program is supported by an NSF/AURA/GMTO risk-reduction program contract to the University of Arizona (GMT-CON-04535, Task Order No. D3 High Contrast Testbed (HCAT), PI Laird Close). The authors acknowledge support from the NSF Cooperative Support award 2013059 under the AURA sub-award NE0651C. Support for this work was also provided by NASA through the NASA Hubble Fellowship grant $\#$HST-HF2-51436.001-A awarded by the Space Telescope Science Institute, which is operated by the Association of Universities for Research in Astronomy Inc. (AURA), under NASA contract NAS5-26555. Maggie Kautz received an NSF Graduate Research Fellowship in 2019. Alex Hedglen received a University of Arizona Graduate and Professional Student Council Research and Project Grant in February 2020. Alex Hedglen and Laird Close were also partially supported by NASA eXoplanet Research Program (XRP) grants 566 80NSSC18K0441 and 80NSSC21K0397 and the Arizona TRIF/University of Arizona “student link” program. We are very grateful for support from the NSF MRI Award $\#$1625441 (for MagAO-X) and funds for the GMagAO-X CoDR from the University of Arizona Space Institute (PI Jared Males) as well. This material is based upon work supported in part by the National Science Foundation as a subaward through Cooperative Agreement AST-1546092 and Cooperative Support Agreement AST-2013059 managed by AURA. The MagAO-X Phase II upgrade program is made possible by the generous support of the Heising-Simons Foundation. We are very grateful for the University of Arizona Space Institute for funding the past CoDR and this PDR effort for GMagAO-X (PI Jared Males). 

\bibliography{General_2} 
\bibliographystyle{spiejour} 

\vspace{2ex}\noindent\textbf{Maggie Kautz} received her Ph.D. from the University of Arizona's James C. Wyant College of Optical Sciences in Tucson, AZ. She is a R$\&$D optical engineer at the Center for Astronomical Adaptive Optics (CAAO) in the Steward Observatory. Her research interests  include optomechanical engineering and optical design for astronomical instrumentation. She received her BS and MS degrees in Optical Engineering and Optical Sciences, respectively, from the University of Arizona.

\vspace{1ex}
\noindent Biographies and photographs of the other authors are not available.


\end{spacing}
\end{document}